\documentstyle[aps]{revtex}
\begin{document}
\twocolumn
\wideabs{
\title{Minimal conditions for the creation of a Friedman--Robertson--Walker 
       universe from a "bounce"}
\author{Carmen Molina--Par\'{\i}s$^{+}$ and Matt Visser$^{++}$}
\address{$^{+}$Theoretical Division, 
         Los Alamos National Laboratory,
         Los Alamos, New Mexico 87545\\
         $^{++}$Physics Department, 
         Washington University,
         Saint Louis, Missouri 63130-4899}
\date{10 September 1998; \LaTeX-ed \today}
\maketitle


{\small {\em Abstract:} In this Letter we investigate the {\em
minimal} conditions under which the creation of our universe might
arise due to a ``bounce'' from a previous collapse, rather than an
explosion from a big-bang singularity. Such a bounce is sometimes
referred to as a {\em Tolman wormhole}. We subject the bounce to a
general model-independent analysis along the lines of that applied to
the Morris--Thorne {\em traversable wormholes}, and show that there is
always an open temporal region surrounding the bounce over which the
strong energy condition (SEC) must be violated. On the other
hand, {\em all } the other energy conditions can easily be
satisfied. In particular, we exhibit an inflation-inspired model in
which a big bounce is ``natural''.  }


} 


Oscillating universes~\cite{Einstein,Tolman} are alternatives to
standard big bang
cosmology~\cite{Hawking-Ellis,Wald,MTW,Peebles}. They avoid the
big-bang singularity and replace it with a cyclical evolution from a
previous incarnation of our present universe. Unfortunately, many of
the older discussions of oscillating universes leave the nature of the
turnaround quite ambiguous (cusp? angular-momentum barrier?).
Interest in oscillating universes largely declined after the
development of the first cosmological singularity
theorem~\cite{Hawking-Ellis,Wald}, but we feel that the time is ripe
for a reassessment of the situation.  In this Letter, we model the
turnaround by a Friedman--Robertson--Walker (FRW) universe undergoing
a ``bounce'' and ask what the absolute minimum requirements are for
such a bounce to occur. Not too surprisingly, the strong energy
condition (SEC) of classical gravity must be
violated~\cite{Bekenstein,Parker-Fulling,Parker-Wang}. (SEC--violation
is a necessary but not sufficient condition.) More surprisingly, for
universes with positive spatial curvature, {\em none} of the other
energy conditions {\em need}\, be violated.  We shall present a
model-independent analysis of the bounce similar to the
model-independent analysis applied to the Morris--Thorne traversable
wormholes
\cite{Morris-Thorne,Visser,Hochberg-Visser,Hochberg-Visser2,Hochberg-Visser3,Hochberg-Visser4},
and also show with specific examples how the various cosmological
singularity theorems~\cite{Hawking-Ellis,Wald} and their modern
extensions~\cite{Borde94a,Borde94b,Borde94c,Borde96,Borde97} can be
evaded.  Finally we discuss the extent to which SEC violations are
compatible with known physics, and exhibit an inflation-inspired model
for which a big-bounce is ``natural''.

\def\implies{\Rightarrow}
\def\sign{\hbox{sign}}

A bouncing baby universe, or Tolman wormhole, is simply a FRW universe
that undergoes a collapse, instant of maximum compression, and
subsequent expansion (as opposed to undergoing a big crunch
singularity or exhibiting a big-bang singularity). In a
model-independent analysis, the key idea is to extract as much
information as possible from the energy conditions without making any
particular commitment to the equation of state for the matter content
of the universe
\cite{Morris-Thorne,Visser,Hochberg-Visser,Hochberg-Visser2,Hochberg-Visser3,Hochberg-Visser4}.
The utility of such an approach has recently been demonstrated in a
different context: applying the energy conditions to the epoch of
galaxy formation~\cite{Galaxy,Galaxy-2,Galaxy-3}.

The FRW cosmology is described by the 
metric~\cite{Hawking-Ellis,Wald,MTW,Peebles}
\begin{equation}
ds^2 = -dt^2 + a(t)^2 
\left[ {dr^2\over1-k r^2} + r^2(d\theta^2+\sin^2\theta d\phi^2)\right],
\end{equation}
with $k = +1$, $0$, or $-1$ for hyperspherical, flat, or hyperbolic
spatial sections, respectively.
%
%

To have a bounce, there must be some time at which the size of the
universe is a minimum. We take this to be time zero, and use a
subscript $*$ to denote quantities evaluated at the bounce $t=0$:
\begin{equation}
\label{E-weak}
\dot a_* = 0;
\qquad
\ddot a_* \geq 0.
\end{equation}
This weak inequality on $\ddot a_*$ is not enough for proving our
deeper results.  Since we want time zero to be a true {\em minimum},
we must have $a>a_*$ for $t\neq0$, with this now being a strict
inequality. An application of the fundamental theorem of differential
calculus now implies:
\begin{equation}
\label{E-strict}
\exists \; \tilde t>0: \forall t \in (-\tilde t,0)\cup(0,\tilde t) 
\qquad \ddot a > 0.
\end{equation}
This is the analog, for a bouncing baby universe, of the
Morris--Thorne {\em ``flare-out''} condition for traversable
wormholes~\cite{Morris-Thorne}. (See also 
\cite[equation~(11.12), page~104]{Visser} and
\cite{Hochberg-Visser,Hochberg-Visser2,Hochberg-Visser3,Hochberg-Visser4}.)
{\em Mutatis mutandis}, there will be similar open regions surrounding
the bounce for which
\begin{equation}
\label{E-strict-2+3}
{d^2[\ln(a)]\over dt^2} >0;
\quad \hbox{and} \quad
{d^2(a^3)\over dt^2} >0, 
\end{equation}
with these again being strict inequalities. 

For a FRW universe the Einstein equations reduce to
\begin{equation}
\rho = {3\over8\pi G}
\left[{\dot a^2\over a^2} + {k\over a^2} \right],
\end{equation}
\begin{equation}
p = -{1\over8\pi G}
\left[2{\ddot a\over a} + {\dot a^2\over a^2} + {k\over a^2} \right].
\end{equation}
Some quantities of interest are,
\begin{eqnarray}
\rho + p 
&=& 
{1\over4\pi G}
\left[ -{\ddot a\over a} + {\dot a^2\over a^2} + {k\over a^2} \right]
\nonumber\\
&=&
{1\over4\pi G}
\left[ -{d^2\over dt^2} \ln a + {k\over a^2} \right],
\end{eqnarray}
\begin{eqnarray}
\rho - p 
&=& 
{1\over4\pi G}
\left[ {\ddot a\over a} + 2{\dot a^2\over a^2} + 2{k\over a^2} \right]
\nonumber\\
&=&
{1\over4\pi G}
\left[ {1\over 3a^3} {d^2 (a^3)\over dt^2}  + 2{k\over a^2} \right],
\end{eqnarray}
and
\begin{equation}
\rho + 3 p = -{3\over4\pi G} \left[{\ddot a\over a} \right].
\end{equation}
By the strict inequalities discussed above [equations
(\ref{E-strict})--(\ref{E-strict-2+3})], there will be open temporal
regions surrounding the bounce for which
\begin{equation}
\rho + p <
{1\over4\pi G}
\left[ {k\over a^2} \right].
\end{equation}
\begin{equation}
\rho - p > 
{1\over4\pi G}
\left[ {k\over a^2} \right].
\end{equation}
\begin{equation}
\rho + 3 p < 0.
\end{equation}


For comparison, the standard point-wise energy conditions are the {\em
null energy condition} (NEC), {\em weak energy condition} (WEC), {\em
strong energy condition} (SEC), and {\em dominant energy condition}
(DEC).  Their specializations to a FRW universe have previously been
discussed in~\cite{Galaxy,Galaxy-2,Galaxy-3}.  Basic definitions are
given in~\cite{Hawking-Ellis,Visser}.
\begin{equation}
\hbox{NEC} \iff \quad 
(\rho + p \geq 0 ).
\end{equation}
\begin{equation}
\hbox{WEC} \iff \quad 
(\rho \geq 0 ) \hbox{ and } (\rho + p \geq 0).
\end{equation}
\begin{equation}
\hbox{SEC} \iff \quad 
(\rho + 3 p \geq 0 ) \hbox{ and } (\rho + p \geq 0).
\end{equation}
\begin{equation}
\hbox{DEC} \iff \quad 
(\rho \geq 0 ) \hbox{ and } (\rho \pm p \geq 0).
\end{equation}
As applied to a bouncing baby universe we have: 
\\ 
First, by working at the bounce itself [where we only have 
the weak inequality (\ref{E-weak})]
\begin{equation}
\exists\ \hbox{Bounce}+[k=-1] \implies \quad 
\hbox{NEC violated},
\end{equation}
\begin{equation}
\exists\ \hbox{Bounce}+[k=0;\ \ddot a_*>0 ] \implies \quad 
\hbox{NEC violated},
\end{equation}
\begin{equation}
\exists\ \hbox{Bounce}+[k=+1;\ \ddot a_* > a_*^{-1}] \implies \quad 
\hbox{NEC violated}.
\end{equation}
(In particular, any of these three conditions automatically implies
violation of the WEC, SEC, and DEC.) Thus a bounce in a hyperbolic
($k=-1$) universe must violate {\em all} the pointwise energy
conditions, a bounce in a spatially flat ($k=0$) universe is on the
verge of violating {\em all} the energy conditions, and a sufficiently
gentle bounce in a hyperspherical ($k=+1$) universe exhibits a
``window of opportunity'' that requires more detailed analysis.
\\
Secondly, by working in suitable open regions surrounding the bounce
[and using the strict inequalities (\ref{E-strict})--(\ref{E-strict-2+3})
derived above] we obtain the stronger results
\begin{equation}
\exists\ \hbox{Bounce} + [k\neq+1] \implies \quad 
\hbox{NEC violated},
\end{equation}
\begin{equation}
\exists\ \hbox{Bounce} \implies \quad \hbox{SEC violated}.
\end{equation}
Thus the energy condition violations are minimized by taking the
universe to be hyperspherical ($k=+1$) and by making the bounce
sufficiently gentle: $\ddot a_* \leq a_*^{-1}$. In this case it is
easy to check that NEC, WEC, and DEC are satisfied, and {\em only}\, SEC
need be violated.  Indeed we only need SEC to be violated in some open
temporal region surrounding the bounce, and it is quite possible to
satisfy {\em all} the point-wise energy conditions at sufficiently
early and late times:
\begin{eqnarray}
&&\exists\ \hbox{Bounce}+[k=+1; \ \ddot a_* \leq a_*^{-1}] \implies 
\nonumber\\
&&\quad 
\hbox{NEC, WEC, DEC satisfied; SEC violated}.
\end{eqnarray}
This is to be contrasted to the situation for Morris--Thorne
traversable wormholes, wherein (for spherically symmetric wormholes)
there is an open spatial region surrounding the throat over which the
NEC (and therefore also the WEC, SEC, and DEC) must be
violated~\cite{Morris-Thorne,Visser,Hochberg-Visser,Hochberg-Visser2}.
Generalization of all these results to spacetimes more general than
the FRW universes, along the lines of
\cite{Hochberg-Visser3,Hochberg-Visser4} is certainly possible, and we
intend to address this issue more fully in a subsequent
paper~\cite{Hochberg-Molina-Visser}.

A simple specific example of a geometry that satisfies NEC, WEC, and
DEC but violates SEC is
\begin{eqnarray}
ds^2 &=& -dt^2 + 
\left( a_*^2 + \beta^2 t^2 \right)
\nonumber\\
&&\times 
\left[ {dr^2\over1-r^2} + r^2 (d\theta^2 + \sin^2\theta\, d\phi^2) \right],
\end{eqnarray}
provided we take $\beta < 1$. Note that this is the temporal analog
of the toy model traversable wormhole considered on page 398 of
the Morris--Thorne article \cite{Morris-Thorne}. Explicit calculation
of the stress-energy components yields
\begin{equation}
\rho = 
{3\over8\pi G} { a_*^2 + \beta^2 [1+ \beta^2] t^2 \over 
\left(a_*^2 + \beta^2 t^2\right)^2 } >0.
\end{equation}
\begin{equation}
p = 
{-1\over8\pi G} { a_*^2 [1+2\beta^2]+ \beta^2 [1+ \beta^2] t^2 \over 
\left(a_*^2 + \beta^2 t^2\right)^2 } <0. 
\end{equation}
So that
\begin{equation}
\rho+p =
{2\over8\pi G}  { a_*^2 [1-\beta^2]+ \beta^2 [1+ \beta^2] t^2 \over 
\left(a_*^2 + \beta^2 t^2\right)^2 } > 0.
\end{equation}
\begin{equation}
\rho-p =
{2\over8\pi G}  { a_*^2 [2+\beta^2]+ 2\beta^2 [1+ \beta^2] t^2 \over 
\left(a_*^2 + \beta^2 t^2\right)^2 } > 0.
\end{equation}
\begin{equation}
\rho+3p =
{- 6\over8\pi G}  { a_*^2 \beta^2  \over 
\left(a_*^2 + \beta^2 t^2\right)^2 } < 0.
\end{equation}
The condition $\beta < 1$ is used to keep $\rho+p\,$ positive definite
and prevent violations of the NEC. The pressure is not positive in
this toy model, (nor does it need to be positive to satisfy the energy
conditions).

A second specific example of a geometry that also satisfies NEC, WEC,
and DEC but violates SEC is
\begin{eqnarray}
ds^2 &=& -dt^2 + 
a_*^2 \cosh^2(H t)
\nonumber\\
&& \times 
\left[ {dr^2\over1-r^2} + r^2 (d\theta^2 + \sin^2\theta\, d\phi^2) \right],
\end{eqnarray}
provided we take $H<a_*^{-1}$. Note that this is {\em not}\, de Sitter
space, since de Sitter space would correspond to $H \equiv a_*^{-1}$.
Explicit calculation of the stress-energy components yields
\begin{equation}
\rho = 
{1\over8\pi G} 
\left( 3 H^2 + { 3 (1 - a_*^2 H^2) \over 
a_*^2 \cosh^2(H t) } \right)
>0.
\end{equation}
\begin{equation}
p = 
{-1\over8\pi G} 
\left( 3 H^2 + { (1 - a_*^2 H^2) \over 
a_*^2 \cosh^2(H t) } \right)
<0.
\end{equation}
So that
\begin{equation}
\rho+p =
{1\over8\pi G}  \left( { 2 (1 - a_*^2 H^2) \over 
a_*^2 \cosh^2(H t) } \right) \geq 0.
\end{equation}
\begin{equation}
\rho-p =
{1\over8\pi G} \left( 6 H^2 +  { 4 (1 - a_*^2 H^2) \over 
a_*^2 \cosh^2(H t) } \right) > 0.
\end{equation}
\begin{equation}
\rho+3p =
{- 6\over8\pi G}  H^2  < 0.
\end{equation}
The condition $H<a_*^{-1}$ is now used to keep $\rho+p$ positive and
prevent violations of the NEC. The pressure is again not positive in
this toy model, nor does it need to be positive to satisfy the energy
conditions.


Perhaps the best known cosmological singularity theorems are the
Penrose-Hawking and Geroch theorems~\cite{Hawking-Ellis,Wald}. Both of
these theorems explicitly use the SEC as an input hypothesis, so
violating the SEC vitiates these theorems. Now when it comes to the
singularity theorems relevant to black hole formation, it was rapidly
realized that the original black hole singularity theorem (which also
uses the SEC~\cite{Hawking-Ellis,Wald}) could be modified to produce
more powerful theorems that used weaker energy conditions (e.g, the
NEC~\cite{Hawking-Ellis,Wald}). It was commonly believed (in at least
some circles) that the cosmological singularity theorems could be
similarly strengthened, and there are in fact a number of newer
cosmological singularity theorems that use the NEC (but at the cost of
adding other rather strong
conditions)~\cite{Borde94a,Borde94b,Borde94c,Borde96,Borde97}. How
does the present discussion evade the consequences of these theorems?

The theorems of~\cite{Borde94a,Borde94b} are stated using the WEC
but really only need the NEC. However the key assumption made there
is that the universe is open in the mathematical sense (which in
a FRW universe implies the universe is either hyperbolic, $k=-1$,
or flat, $k=0$). Thus these singularity theorems are compatible
with the results of this Letter since we have explicitly shown that
a bounce in hyperbolic or flat FRW universes requires NEC violation.

The closed universe singularity theorem of~\cite{Borde94c} uses a very
strong technical requirement (compact localized past light cones)
explicitly violated by our models. The more recent results
in~\cite{Borde96,Borde97} are also compatible with our results in that
the theorems either apply to open universes, or make additional
technical assumptions violated by our analysis.


{\em Physical reasonableness of the SEC:} It is relatively difficult
to violate the NEC, WEC, and DEC; violations of these energy
conditions typically being due to (small) quantum effects. On the
other hand, it is rather easy to violate the SEC, leading some
researchers to refer to the SEC as ``the unphysical energy condition''.
Violations of the SEC are generic to classical scalar
fields~\cite{Hawking-Ellis,Visser}, to inflationary
spacetimes~\cite{Borde94a,Borde94b,Borde94c,Borde96,Borde97}, to
spacetimes with (cosmologically large) positive cosmological
constants~\cite{Galaxy,Galaxy-2,Galaxy-3}, to certain mean-field
quantum field theories~\cite{Rose86,Rose87} and other quantum
mechanical situations~\cite{Parker-Fulling}, and significantly, to
classical relativistic fluids with two body
interactions~\cite{Parker-Wang,Kandrup}. A particularly wide class
of Tolman wormholes can also be constructed by Wick rotating Euclidean
wormholes back to Lorentzian signature~\cite{Coule}; the Wick
rotated Euclidean wormholes typically satisfying all energy conditions
except the SEC. These observations are important in that they
elevate the discussion of this Letter from a mere mathematical
curiosity to an issue that merits serious attention.


{\em An inflation-inspired model:} The generic feature common to all
inflationary FRW models is the introduction of a minimally coupled
scalar field $\phi$ called the inflaton (in addition to whatever
matter is normally present). The inflaton contributes to the energy
density and pressure:
\begin{equation}
\rho_\phi = {1\over2} \dot \phi^2 + V(\phi),
\end{equation}
\begin{equation}
p_\phi  = {1\over2} \dot \phi^2 - V(\phi),
\end{equation}
and the inflaton field satisfies the equation of motion
\begin{equation}
\ddot \phi + 3 {\dot a \over a} \dot \phi 
+ {\partial V(\phi)\over\partial\phi} = 0.
\end{equation}
The key feature relevant for the present discussion is that
\begin{equation}
\rho_\phi + 3 p_\phi =  2 \dot \phi^2 - 2 V(\phi).
\end{equation}
If the inflaton field bounces at the same time as the geometry ({\em
i.e.}, $\dot\phi_*=0$), then the inflaton field can be used as the
natural candidate for providing the SEC violations required to support
the bounce:
\begin{equation}
(\rho+3p)_{\mathrm total} = 
(\rho+3p)_{\mathrm normal} +  2 \dot \phi^2 - 2 V(\phi).
\end{equation}
\begin{equation}
(\rho+p)_{\mathrm total} = 
(\rho+p)_{\mathrm normal} + \dot \phi^2.
\end{equation}
\begin{equation}
(\rho-p)_{\mathrm total} = 
(\rho-p)_{\mathrm normal} +  2 V(\phi).
\end{equation}
Thus adding a spatially-homogeneous inflaton field to normal (energy
condition satisfying) matter preserves the NEC, WEC, and DEC, but can
easily lead to violations of the SEC.  In this sense inflation (either
old or new inflation; but not chaotic or eternal inflation) is
naturally compatible with the bounce scenario. With typical estimates
for the inflaton VEV being of order the GUT energy scale, we would
similarly estimate the bounce to occur when the radius of the universe
is about one GUT distance scale (about 1~000 Planck lengths). This is
certainly a small distance, even by particle physics standards, but
because it is so much larger than the Planck scale, we may still
reasonably hope for the applicability of semiclassical quantum gravity
--- thus we now hold out the reasonable hope for a big bounce that not
only evades the classical singularity theorems but also evades the
necessity for dealing with the full theory of quantum gravity.

In summary, (1) replacing the big bang with a big bounce violates the
SEC but does not necessarily violate any of the other energy
conditions, and (2) violating the SEC is relatively easy and can be
achieved at the classical level, without needing to appeal to quantum
effects. Of course, even more exotic variations can be
contemplated. You can consider the effects of violating {\em all} the
energy conditions \cite{Levin}, including the NEC, or even more boldly
you can consider having the universe bootstrap itself into existence
via a chronology violating region \cite{Gott-Li}. A key aspect of this
Letter is that extreme steps of this type are not necessary: the big
bang singularity can be tamed with relatively mild modifications of
the standard cosmological model. Indeed, there are simple extensions
of either old or new inflation for which such a bounce is ``natural''.

\bigskip
  

{\em Acknowledgments:} This research was supported by the
U.S. Department of Energy.  Matt Visser also wishes to thank LAEFF
(Laboratorio de Astrof\'\i{}sica Espacial y F\'\i{}sica Fundamental;
Madrid, Spain), and Victoria University (Te Whare Wananga o te Upoko
o te Ika a Maui; Wellington, New Zealand) for hospitality during final
stages of this work.


\end{document}